\documentclass[aps,prl,amsmath,twocolumn,floatfix,superscriptaddress,longbibliography,footinbib]{revtex4-1}
\usepackage{amsmath, amsfonts, amssymb, bm, bbm}
\usepackage{mathtools}
\usepackage[inline]{enumitem}
\usepackage{graphicx}
\usepackage{xcolor}
\usepackage[breaklinks]{hyperref}
\hypersetup{
	pdfnewwindow=true,      
	colorlinks=true,       
	linkcolor=blue,          
	citecolor=magenta,        
	filecolor=blue,      
	urlcolor=blue        
}

\usepackage{sidecap,tikz}
\definecolor{lime}{HTML}{A6CE39}
\DeclareRobustCommand{\orcidicon}{\hspace{-1mm}
	\begin{tikzpicture}
		\draw[lime, fill=lime] (0,0) 
		circle [radius=0.16] 
		node[white] {{\fontfamily{qag}\selectfont \tiny \,ID}};
		\draw[white, fill=white] (-0.0525,0.095) 
		circle [radius=0.007];
	\end{tikzpicture}
	\hspace{-3mm}
}
\foreach \x in {A, ..., Z}{\expandafter\xdef\csname orcid\x\endcsname{\noexpand\href{https://orcid.org/\csname orcidauthor\x\endcsname}
		{\noexpand\orcidicon}}
}
\usepackage[capitalize]{cleveref} 
\usepackage{glossaries}
\glsdisablehyper 
\newacronym{wtd}{WTD}{waiting time distribution}
\newacronym{mbs}{MBS}{Majorana bound state}
\newacronym{qshi}{QSHI}{quantum spin Hall insulator}
\newcommand{\up}{\uparrow}
\newcommand{\dw}{\downarrow}
\newcommand{\e}{\mathrm{e}}

\DeclarePairedDelimiter\mean{\langle}{\rangle}

\newcommand{\bea}{\begin{eqnarray}}
	\newcommand{\eea}{\end{eqnarray}}
\newcommand{\beq}{\begin{equation}}  
	\newcommand{\eeq}{\end{equation}}
\newcommand{\non}{\nonumber}

\begin{document} 

\title{Nonlocality of Majorana bound states revealed by electron waiting times in a topological Andreev interferometer}

\author{Paramita Dutta\orcidA{}}
\affiliation{Theoretical Physics Division, Physical Research Laboratory, Ahmedabad-380009, India}

\author{Jorge Cayao\orcidB{}}
\affiliation{Department of Physics and Astronomy, Uppsala University, Box 516, S-751 20 Uppsala, Sweden}

\author{Annica M. Black-Schaffer\orcidC{}}
\affiliation{Department of Physics and Astronomy, Uppsala University, Box 516, S-751 20 Uppsala, Sweden}

\author{Pablo Burset\orcidD{}}
\affiliation{Department of Theoretical Condensed Matter Physics, Condensed Matter Physics Center (IFIMAC) and Instituto Nicol\'as Cabrera, Universidad Aut\'onoma de Madrid, 28049 Madrid, Spain}

\date \today

\begin{abstract}
The analysis of waiting times of electron transfers has recently become experimentally accessible owing to advances in noninvasive probes working in the short-time regime. We study  electron waiting times in a topological Andreev interferometer: a superconducting loop with controllable phase difference connected to a quantum spin Hall edge, where the edge state helicity enables the transfer of electrons and holes into separate leads, with transmission controlled by the loop's phase difference $\phi$. This setup features gapless Majorana bound states at $\phi=\pi$. The waiting times for electron transfers across the junction are sensitive to the presence of the gapless states, but are uncorrelated for all $\phi$. By contrast, at $\phi=\pi$ the waiting times of Andreev-scattered holes show a strong correlation and the crossed (hole-electron) distributions feature a unique behavior. Both effects exclusively result from the nonlocal properties of Majorana bound states. Consequently, electron waiting times and their correlations could circumvent some of the challenges for detecting topological superconductivity and Majorana states beyond conductance signatures. 
\end{abstract}

\maketitle

Fluctuations in electron transport can greatly impact the performance of electronic circuits but, at the same time, provide us with invaluable information about the quantum-coherent behavior of conductors~\cite{Blanter2000}. 
Charge fluctuations are often analyzed by full counting statistics~\cite{Levitov1996,Bagrets2003,Bagrets2006}, which usually concerns the zero-frequency or long-time limit. However, experimental advances with noninvasive probes~\cite{Camus2017,Geller2019,Pekola2019} have now enabled access to the short-time regime with almost single-event resolution. 

A prominent, recently developed tool for the short-time regime is the electron waiting time distribution (WTD): the distribution of time intervals between consecutive charge transfers. Electron WTDs provide information about tunneling events in mesoscopic conductors beyond average current and noise~\cite{Albert2011,Albert2012,Haack2014,Tang2014,Thomas2014,Sothman2014,Dasenbrook2015,Burset2019,Kosov2019,Kosov2021}, and have recently been measured in quantum dots connected to metallic electrodes~\cite{Gorman2017,Geller2019,Tarucha2020,Brange2021} and superconductors~\cite{Jenei2019,Ranni2021}.  
Moreover, correlations between waiting times indicate nonrenewal quantum transport~\cite{Kosov2019}, where electron transfers are not independent or identically distributed, making WTDs a source of information distinct from other statistical tools~\cite{Dasenbrook2015,Kosov2019,Kosov2021}.

\begin{figure}[hb!]
	\centering
	\includegraphics[scale=0.65]{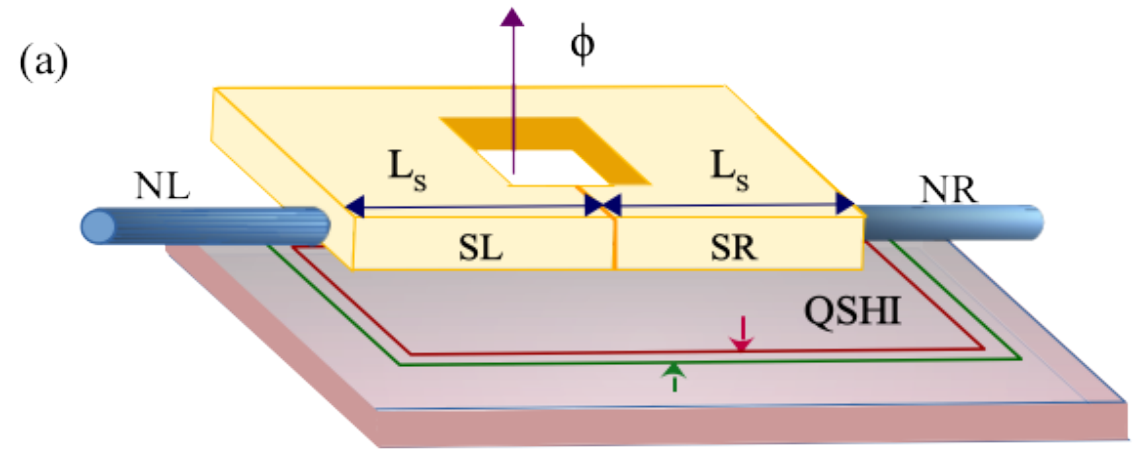}
	\includegraphics[scale=0.16]{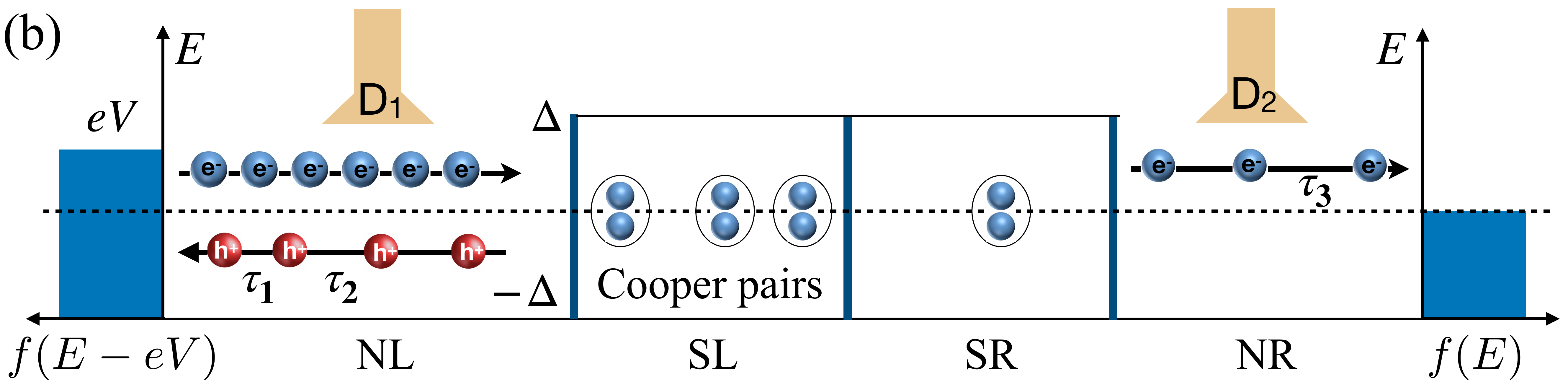}
	\caption{
		(a) Andreev interferometer at the QSHI edge with magnetic flux $\phi$ applied through a superconductor loop. 
		(b) Incident electrons (blue balls) with energy within the gap $\Delta$ scatter off the NL-SL-SR-NR junction only as electron transmissions to NR or Andreev-reflected holes (red balls) into NL. Detectors D$_1$ and D$_2$, respectively placed at NL and NR, detect electrons or holes either individually or simultaneously. NL is biased by a voltage $V$ and NR is at equilibrium. }
	\label{fig:model}
\end{figure}

In this Letter, we analyze the potential of electron WTDs and their correlations for identifying topological superconductors hosting Majorana bound states (MBSs)~\cite{Kane2005,Fu2008,Fu2009,Qi2011}. 
There is currently an intense research activity focused on obtaining reliable signatures of MBSs, since the simplest one, a robust and quantized zero-bias conductance peak~\cite{Mourik2012,Higginbotham2015,Deng2016}, has proven insufficient~\cite{Frolov2019,Frolov2021}. Electron waiting times in superconducting hybrid junctions have already been proposed to characterize the entanglement between the electrons in Cooper pairs~\cite{Rajabi2013,Albert2016,Walldorf2018} and to detect the presence of MBSs~\cite{Chevallier2016,Mi2018,Weymann2020,Fu2022,Schulz2022}. 
These theoretical proposals extended the concept of waiting times to both spin and electron-hole degrees of freedom, but they have still primarily only focused on the \textit{local} properties of MBSs~\cite{Chevallier2016,Mi2018,Weymann2020,Fu2022,Schulz2022}. 
Instead, we here suggest a Majorana platform without magnetic materials that is both conceptually simple and presents important advantages for measuring waiting times of electrons and holes and their \textit{nonlocal} properties: 
an Andreev interferometer built on the edge of a quantum spin Hall insulator (QSHI)~\cite{Konig2007,Brune2010,Sullivan2011,Brune2012,Reis2017,Kammhuber2017,Wu2018} [\cref{fig:model}(a)]. 

The QSHI features helical edge states consisting of one-dimensional Dirac fermions characterized by spin-momentum locking~\cite{Wu2006,Xu2006}. When proximitized by a narrow superconducting lead~\cite{Sullivan2012,Hart2014,Wiedenmann2016,Bocquillon2017,Deacon2017,Hart2017,Sajadi2018,Fatemi2018}, the helical edge states guarantee that only electrons tunnel through the lead and only Andreev-converted holes are reflected~\cite{Adroguer2010,Black-Schaffer2012,Black-Schaffer2013,Crepin2015,Tkachov2015,Cayao2017,Keidel2018,Keidel2020,Cayao2022,Lu2022}. 
Consequently, the superconductor acts as a beam splitter that separates electrons from holes into different leads and allows independent detection, see \cref{fig:model}(b). 
In the interferometer setup~\cite{Chandrasekhar1998,Chandrasekhar2001}, a superconducting loop with controllable phase difference $\phi$ is connected to the QSHI edge [\cref{fig:model}(a)], so that the electric conductance at the NL or NR sides depends periodically on $\phi$. Importantly, recent experiments~\cite{Wiedenmann2016,Bocquillon2017,Deacon2017} have found that the lowest energy bound states formed at the SL-SR interface are always gapless at $\phi=\pi$, and thus MBSs~\cite{Fu2008,Fu2009}. 

Due to the lack of a gap, the topological MBSs dominate the local and nonlocal transport across the interferometer. 
We find that the waiting times for electron transfers across our junction are sensitive to the MBSs, but are uncorrelated with each other. 
By contrast, the waiting times of Andreev reflected holes are less sensitive to the MBSs, but instead present a strong correlation at $\phi\sim\pi$. 
Importantly, the crossed (hole-electron) distributions and their correlations feature a unique behavior characteristic of a gapless nonlocal MBS. 
Consequently, electron waiting times and their correlations constitute an alternative signature of MBSs, sensitive to their nonlocal nature, thus circumventing the problems arising from trivial resonant levels that naturally form in many Majorana platforms~\cite{Frolov2019}. 

\textit{Topological Andreev interferometer.---}
We consider an Andreev interferometer at the edge of a QSHI (\cref{fig:model}), which comprises of a superconducting loop with a short SL-SR junction that is attached to the normal metal leads NL and NR. 
For simplicity, we fix the length of each superconductor segment to be equal, $L_{\text S}$, and only analyze the situation where SL and SR share an interface~[\cref{fig:model}(b)]~\footnotemark[1]\footnotetext{A finite separation of the SL and SR arms, or asymmetry in the superconductor lengths, would not qualitatively affect our results.}. Low-energy excitations are described in the basis $\Psi(x)=(
\psi_{\uparrow},
\psi_{\downarrow},
\psi_{\downarrow}^{\dagger},
-\psi_{\uparrow}^{\dagger})^{T}$, with $\psi^{\dagger}_{\sigma}(x)$ the creation operator for electrons with spin $\sigma\in\{\up,\dw\}$ at position $x$, by the Bogoliubov-de Gennes Hamiltonian~\cite{Cayao2017} 
\begin{equation}\label{Eq:BdG_H}
H_{\rm BdG}=\hbar v_{F}k_{x} \hat{\eta}_{3}\hat{\sigma}_{3}-\mu \hat{\eta}_{3}\hat{\sigma}_{0} + \Delta (x) \hat{\eta}_{1} \hat{\sigma}_{0} .
\end{equation}
Here, $v_{F}$ is the Fermi velocity, $\mu$ the chemical potential, and the Pauli matrices $\hat{\eta}_{j}$ and $\hat{\sigma}_{j}$ act in Nambu and spin spaces, respectively. 
We set the pair potential $\Delta(x)\!=\!\Delta$ for SL, $\Delta(x)\!=\!\Delta e^{i \phi}$ for SR ($\phi$ is the superconducting phase difference), and zero otherwise. 
Henceforth, 
we set $v_F\!=\!\hbar\!=\!\Delta\!=\!1$ so that the superconducting coherence length is $\xi\!=\!\hbar v_F/\Delta\!=\!1$~\footnotemark[2], 
\footnotetext{Superconductor-normal metal-superconductor junctions with the normal intermediate region much smaller than the superconducting coherence length are usually referred to as \textit{short} junctions. We work in this regime, but use the terms \textit{short} and \textit{long} junctions to distinguish the cases where the total size of the SL-SR segment ($2L_\text{S}$) is smaller or larger, respectively, than the superconducting coherence length $\xi$. }
and set $\mu\!=\!0$ and $eV=\Delta/2$~\footnotemark[3], 
\footnotetext{Our results are mostly insensitive to the chemical potential and do not qualitatively change for voltages within the gap. }
see Supplementary Material (SM)~\cite{SM}. 
The QSHI Andreev interferometer forms a topological Josephson-like junction that hosts gapless MBSs at the SL-SR interface at $\phi=\pi$~\cite{Fu2009,Cayao2022}. For any other trivial junction, the bound states develop a gap around $\phi\sim\pi$. To distinguish between topological (gapless) and trivial (gapful) bound states, we compare below the prototypical cases $\phi=0$ and $\phi=\pi$. Although our junction is always topological, results at $\phi=0$ are qualitatively equivalent to those of a trivial bound state (for any $\phi$), as long as its gap is comparable or larger than the bias $eV$.

\textit{Electron waiting times.---}
WTDs for phase-coherent transport of noninteracting electrons are evaluated from the scattering matrix~\cite{Albert2012,Haack2014,Mi2018}. 
Generally an Andreev interferometer has four effective transport channels [electrons or holes (e, h), incoming or outgoing (i, o), from the left or right leads (L, R)], represented by the spinor $\Psi^{i(o)}=(\psi_{eL}^{i,(o)}, \psi_{hL}^{i(o)}, \psi_{eR}^{i(o)}, \psi_{hR}^{i(o)})^T$. For a given energy $E$, the scattering matrix connects outgoing and incoming solutions of \cref{Eq:BdG_H} as $\Psi^{(o)}= \mathcal{S}\Psi^{(i)}$~\cite{SM}. Owing to the spin-momentum locking at the QSHI edge, here only the normal transmissions $\mathcal{S}_{\alpha L,\alpha R}$ and $ \mathcal{S}_{\alpha R,\alpha L}$, with $\alpha=e,h$, and the Andreev reflections $ \mathcal{S}_{e X, hX}$ and $ \mathcal{S}_{h X, e X}$, with $X=L,R$, are nonzero. From the scattering matrix we define the idle-time probability $\Pi(\{\tau_\gamma\})$ that no particles of type $\gamma=\alpha X$ are detected during the time interval $\tau_\gamma$ (for the stationary processes considered here only time intervals are relevant). Following Refs.~\cite{Haack2014, Mi2018}, we have
\begin{equation}\label{eq:itp}
	\Pi(\{\tau_\gamma\})= \mathrm{det}[\mathcal{I}-\mathcal{S}^{\dagger}(E) \mathcal{K}(\{\tau_\gamma\},E-E') \mathcal{S}(E') ] , 
\end{equation}
where $\mathcal{I}$ is the identity matrix and $\mathcal{K}$ is a diagonal matrix, 
$\mathcal{K}(\{\tau_\gamma\},E) =\bigoplus_{\gamma} K( \tau_\gamma, E)$,
with the kernels~\cite{SM}
\begin{equation}\label{eq:kernel}
	K(\tau_\gamma, E)= \kappa \e^{-i E \tau_\gamma/2 } \sin (E \tau_\gamma/2) / (\pi E ) . 
\end{equation}
The linear dispersion relation of the QSHI helical edge states allows us to naturally divide the transport window $[\mu , \mu + eV ]$ in intervals of width $\kappa=eV/N$, where $N$ is the total number of intervals and $eV$ the applied bias. Due to the inversion symmetry of our setup, we only consider voltages applied to the left lead, $V_L\equiv V$, $V_R=0$. We work in the limit $N\rightarrow\infty$ where \cref{eq:kernel} correctly applies to stationary transport~\cite{Haack2014}. 

We can now define $\mathcal{W}_{\alpha\beta}(\tau)= -\mean{\tau_\alpha}\partial_\tau^2 \Pi(\tau)$ as the probability density of detecting a particle of type $\beta$ at a time $\tau$ after having measured a particle of type $\alpha$~\cite{SM}. Here, the mean waiting time $\mean{\tau_\alpha}$ is related to the average current for $\alpha$ particles, $I_\alpha=1/\mean{\tau_\alpha}$. 
Analogously, we define the joint waiting time $\mathcal{W}_{\alpha\gamma\beta}(\tau_1,\tau_2)= \mean{\tau_\alpha} \partial_{\tau_1} \partial_{\tau_2}^2\Pi$, which generalizes the waiting time distribution between particles of type $\alpha$ and $\beta$ to include the extra detection of a particle of type $\gamma$ at an intermediate time $\tau_1$, such that $0 \leq \tau_1 \leq \tau_2$~\cite{SM}. 
The joint WTD describes correlations between consecutive waiting times. When the waiting times are uncorrelated, the joint distribution factorizes as the product of two waiting time distributions~\cite{Dasenbrook2015}, $\mathcal{W}^\text{unc}_{\alpha\gamma\beta}(\tau_1,\tau_2)= \mathcal{W}_{\alpha\gamma}(\tau_1)\mathcal{W}_{\gamma\beta}(\tau_2)$. We can further quantify the correlations between consecutive waiting times using the correlation function
\begin{equation}\label{eq:correlations}
	\delta\mathcal{W}_{\alpha\gamma\beta}(\tau_1,\tau_2) = \frac{\mathcal{W}_{\alpha\gamma\beta}(\tau_1,\tau_2) - \mathcal{W}_{\alpha\gamma}(\tau_1)\mathcal{W}_{\gamma\beta}(\tau_2)}{\mathcal{W}_{\alpha\gamma}(\tau_1)\mathcal{W}_{\gamma\beta}(\tau_2)} .
\end{equation}

A main feature of the QSHI topological Andreev interferometer is that for an electron (say, spin-up) injected in NL, only (spin-down) holes and (spin-up) electrons can scatter into NL and NR, respectively. Thus, with electrons and holes always scattering into different leads, all local or same detector WTDs are necessarily given by $\mathcal{W}_{ee}$ and $\mathcal{W}_{hh}$, while all \textit{nonlocal} WTDs are given by $\mathcal{W}_{eh}$ and $\mathcal{W}_{he}$, where measurements take place at different detectors. Similarly, $\mathcal{W}_{eee}$ and $\mathcal{W}_{hhh}$ are local joint WTDs, while joint distributions combining electron (NR) and hole (NL) measurements, like $\mathcal{W}_{ehe}$, are nonlocal.
\begin{figure}[t!]
	\centering
	\includegraphics[scale=0.5]{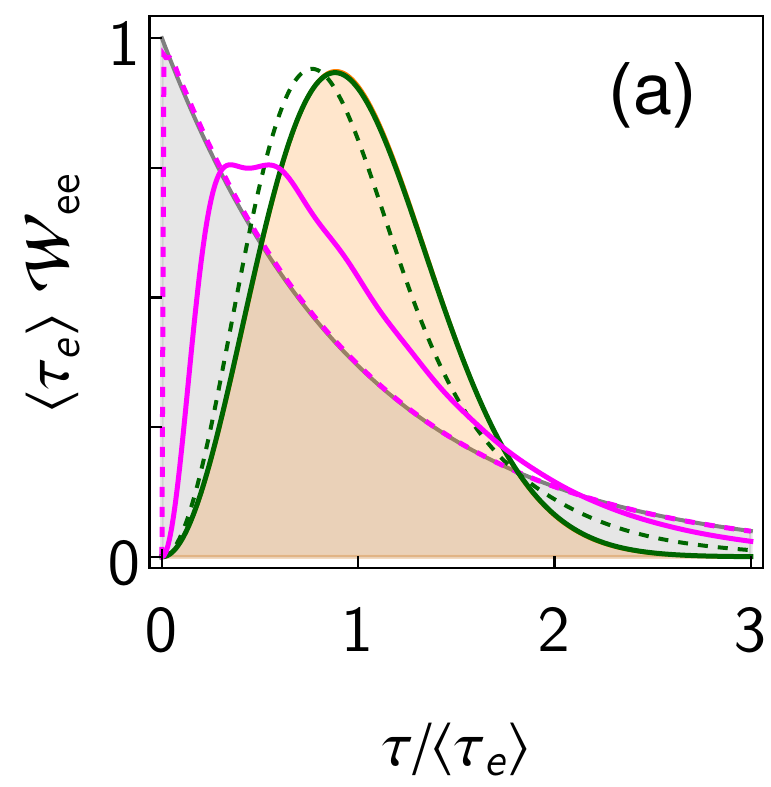}
	\includegraphics[scale=0.49]{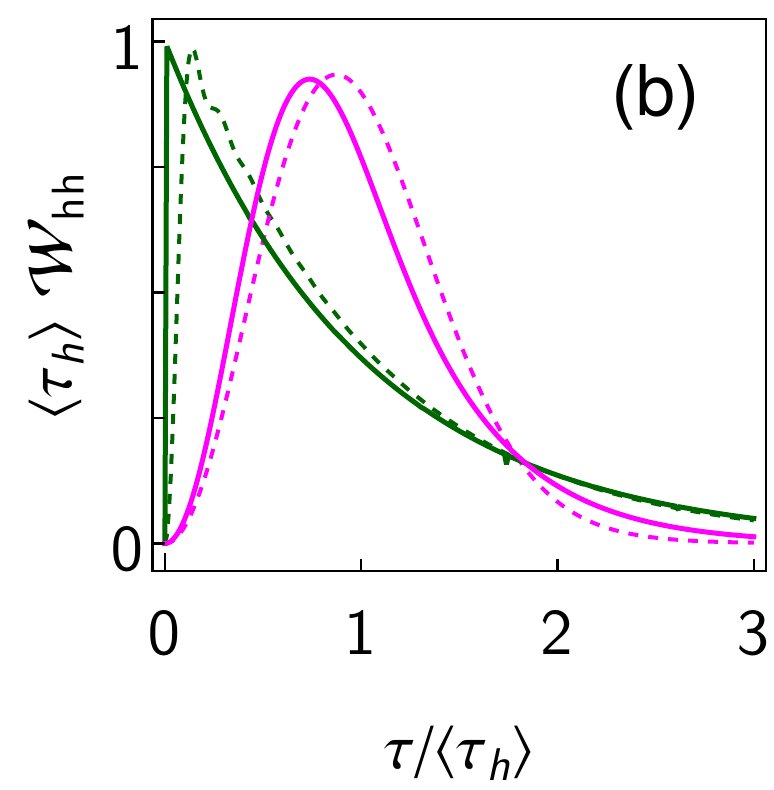}
	\caption{Local WTDs, $\mathcal{W}_{ee}$ (a) and $\mathcal{W}_{hh}$ (b), for $L_{\rm S}=0.2$ (short junction, green) and $L_{\rm S}=1.5$ (long junction, magenta), with $\phi\! =\!0$ for dashed lines and $\phi\! =\!\pi$ for solid ones. 
	The gray (orange) shaded region indicate the Poisson (Wigner-Dyson) distribution. }
	\label{fig:localWTD}
\end{figure}

\textit{Local waiting times.---} We start with the local WTDs $\mathcal{W}_{\alpha\alpha}$, representing two consecutive detections at either NL ($\alpha=h$) or NR ($\alpha=e$). 
%
It was established in Ref.~[\onlinecite{Haack2014}] that the WTD of a quantum-coherent channel with energy-independent transmission is determined by its scattering probability: highly-transmitting channels result in a Wigner-Dyson distribution [orange area in \cref{fig:localWTD}(a)], describing a coherent particle flow, while low-transmitting channels result in a Poisson distribution [gray area in \cref{fig:localWTD}(a)], characteristic of tunnel transport. In both cases, the impossibility of a simultaneous measurement of two particles at the same detector due to the Pauli exclusion principle forces the WTDs to be zero at $\tau=0$. 
Owing to the constrained transport at the QSHI edge, with, e.g., $|\mathcal{S}_{hL,eL}|^2+|\mathcal{S}_{eR,eL}|^2=1$, the transition between the Wigner-Dyson and Poisson distributions is here controlled by the length of the Andreev interferometer, $2L_\text{S}$. Long (short) junctions~\footnotemark[2] with $L_\text{S}>\xi$ ($L_\text{S}<\xi$) give a high probability of Andreev reflection $|\mathcal{S}_{hL,eL}|^2$ (electron transmission $|\mathcal{S}_{eR,eL}|^2$), and result in a Poisson (Wigner-Dyson) distribution~\cite{SM}. This behavior is consistent with earlier results~\cite{Haack2014,Mi2018,Schulz2022}, since the scattering probabilities for the topological Andreev interferometer are almost constant at subgap energies~\cite{SM}.
As a result, any gapped state, at any phase $\phi$, qualitatively follows the $\phi=0$ local WTDs represented in \cref{fig:localWTD} by dashed lines. 

By contrast, for gapless MBSs around $\phi\sim\pi$ the low-energy electron transmission probability becomes strongly energy-dependent for long junctions due to the resonant-tunneling through the MBS~\cite{SM,Cayao2022}. Consequently, we find that $\mathcal{W}_{ee}$ converges to the WTD of a resonant level in the tunnel limit~\cite{Haack2014}, [solid magenta line in \cref{fig:localWTD}(a)], instead of evolving into a Poisson distribution like for $\phi=0$. 
At high transmission (short junctions), the variation with the phase of $\mathcal{W}_{ee}$ is less noticeable [green lines in \cref{fig:localWTD}(a)]. This is also the case for the distribution of reflected holes $\mathcal{W}_{hh}$ at any transparency [\cref{fig:localWTD}(b)], since the probability of Andreev reflection is $|\mathcal{S}_{hL,eL}|^2\sim1$ for all the energies in the transport window ($|E|\leq eV$). 

\begin{figure}[t!]
	\centering
	\includegraphics[scale=0.32]{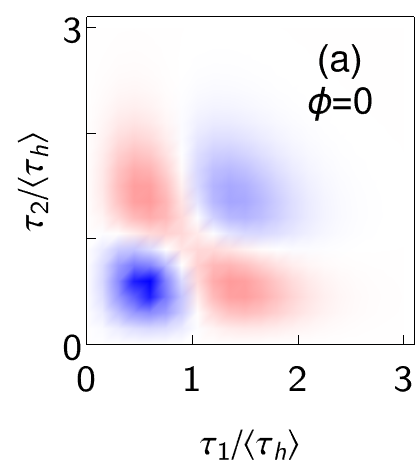}
	\includegraphics[scale=0.32]{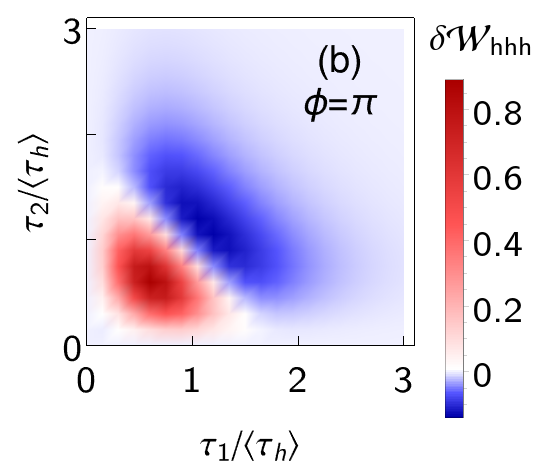}
	\caption{Correlations between waiting times for three consecutive hole detections at NL, $\delta\mathcal{W}_{hhh}(\tau_1,\tau_2)$, for $L_{\rm S}\!=\!1.5$ as a function of $\tau_1$ and $\tau_2$ at (a) $\phi\!=\!0$ and (b) $\phi\!=\!\pi$. }
	\label{fig:jointWTD}
\end{figure}
 
Even though $\mathcal{W}_{hh}$ is not very sensitive to the presence of the MBS, the correlations between consecutive waiting times for hole transfers, $\delta\mathcal{W}_{hhh}$, contain very relevant information. We focus on long junctions, $L_\text{S}>\xi$, which are dominated by Andreev reflection processes and feature a more pronounced dependence on the phase $\phi$. For gapful states ($\phi=0$), the Andreev interferometer behaves like an electron-hole beam splitter, featuring the same correlations as a standard quantum point contact for electrons~\cite{Dasenbrook2015} [\cref{fig:jointWTD}(a)]: 
When the time between two hole transfers is small, $\tau_1<\mean{\tau_h}$ (or long, $\tau_1>\mean{\tau_h}$), the next hole detection at $\tau_2$ will require a long (short) waiting time (red color signals positive correlations). 
By contrast, gapless MBSs around $\phi\sim\pi$ exhibit correlations that seemingly explode at short waiting times, with values increasing an order of magnitude compared to the gapful case [\cref{fig:jointWTD}(b)]. This means that short time intervals between detections are the most likely. The waiting times between electron transfers, on the other hand, are completely uncorrelated, i.e., $\delta\mathcal{W}_{eee}(\tau_1,\tau_2)\simeq \mathcal{W}_{ee}(\tau_1)\mathcal{W}_{ee}(\tau_2)$~\cite{SM}. 

\begin{figure}[t!]
	\centering
	\includegraphics[scale=0.5]{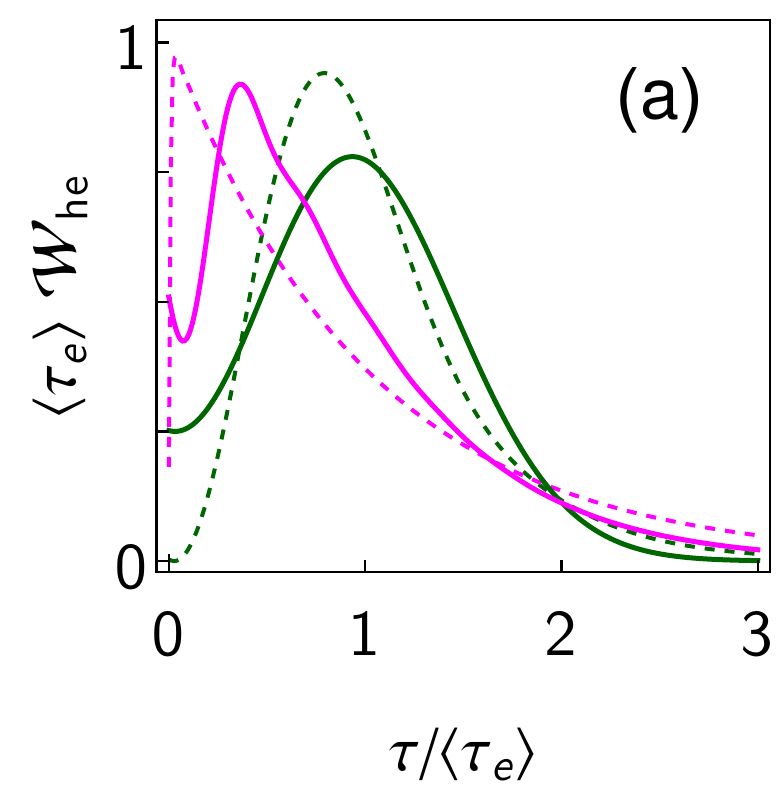}
	\includegraphics[scale=0.5]{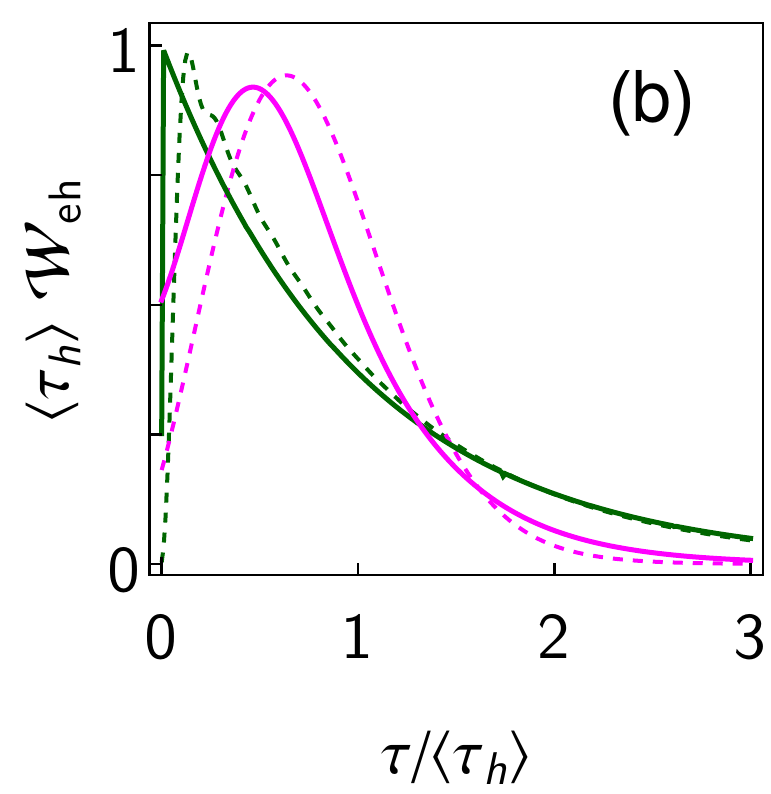}
	\caption{Nonlocal WTDs, $\mathcal{W}_{he}$ (a) and $\mathcal{W}_{eh}$ (b), for $L_{\rm S}\!=\!0.2$ (short junction, green) and $L_{\rm S}=1.5$ (long junction, magenta), with $\phi\! =\!0$ for dashed lines and $\phi\! =\!\pi$ for solid ones. }
	\label{fig:nonlocalWTD}
\end{figure}

\textit{Nonlocal waiting times.---} 
We now fully exploit the multi-terminal advantage of the topological Andreev interferometer by exploring the nonlocal WTDs. By definition, the distribution $\mathcal{W}_{\alpha\beta}$, with $\beta\neq\alpha$, assumes that the first particle $\alpha$ has been detected; no matter how unlikely that event is. Therefore, the nonlocal waiting times are determined by the probability of the second detection. Consequently, $\mathcal{W}_{eh}$ ($\mathcal{W}_{he}$) is determined by the Andreev reflection (electron transmission) probability, following a behavior similar to $\mathcal{W}_{hh}$ ($\mathcal{W}_{ee}$), which we verify in \cref{fig:nonlocalWTD}. The one marked difference between local and nonlocal distributions is that, as particles transfer into different detectors, nonlocal WTDs can be finite at zero waiting time and also always fulfill $\mathcal{W}_{eh}(0)=\mathcal{W}_{he}(0)$~\cite{Mi2018}. 

The nonlocal WTDs at zero waiting time have already been established to increase in the presence of MBSs, independently of the scattering probabilities~\cite{Mi2018}. Here, we interestingly also find that $\mathcal{W}_{he}(\tau)$, which for $\phi=\pi$ is determined by the Majorana-assisted electron tunneling, is further strongly altered. Specifically, $\mathcal{W}_{he}(\tau)$ presents a dip at short but finite waiting times. We explain this behavior as being due to the transition between a regime dominated by the Andreev reflection probability at $\tau\rightarrow0$ into a regime where electron transmissions dominate at long waiting times. The former initially reduces the probability, while the latter imposes a behavior similar to the local distribution, $\mathcal{W}_{ee}$. The dip, or local minimum at short waiting times, reflects this transition and is particularly visible in the presence of Majorana-induced resonant tunneling when the corresponding local WTD becomes anomalous, see \cref{fig:localWTD}(a). With $\mathcal{W}_{eh}$ being primarily determined by $\mathcal{W}_{hh}$, we find no such dip in $\mathcal{W}_{eh}$. This behavior of $\mathcal{W}_{he}$ is unique to the topological Andreev interferometer, which we confirmed by checking both local and nonlocal WTDs for an ordinary interferometer, in the absence of any topology. 

\begin{figure}[t!]
	\centering
	\includegraphics[scale=0.331]{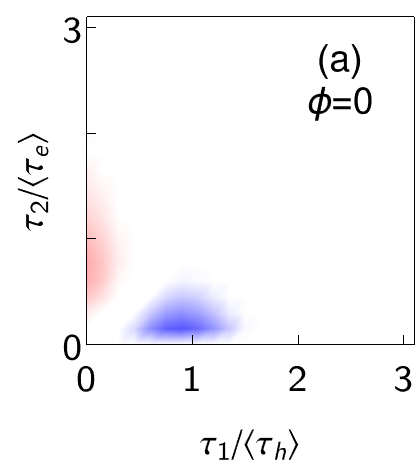}
	\includegraphics[scale=0.327]{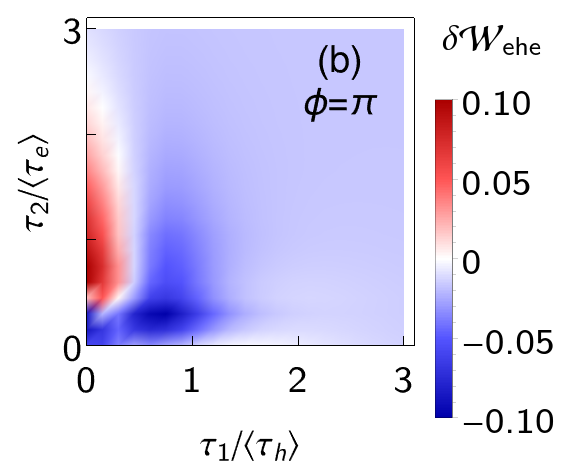}
	\caption{Correlations between waiting times for the detection sequence electron-hole-electron between NL and NR: $\delta\mathcal{W}_{ehe}(\tau_1,\tau_2)$ for $L_{\rm S}\!=\!1.5$ as a function of $\tau_1$ and $\tau_2$ at (a) $\phi\!=\!0$ and (b) $\phi\!=\!\pi$. }
	\label{fig:jointWTD2}
\end{figure}

We further find that the correlations between nonlocal waiting times also show a unique dependence on the phase $\phi$. 
We focus on alternate electron-hole-electron transfers in the long junction regime, where the phase dependence is stronger, and study the behavior of $\mathcal{W}_{ehe}$ containing the correlations between $\mathcal{W}_{he}$ and $\mathcal{W}_{eh}$. Very short waiting times, $\tau_1<\mean{\tau_{h}}$, show only a weak correlation with long waiting times ($\tau_2>\mean{\tau_{e}}$) for gapful states ($\phi=0$), but this behavior is enhanced two orders of magnitude for gapless MBS around $\phi\sim\pi$ [\cref{fig:jointWTD2}]. As mentioned above, the sequential tunneling of three electrons, $\mathcal{W}_{eee}$, is uncorrelated. However, including one hole detection between the electron transfers drastically changes the statistics in the presence of MBSs. These results indicate that to completely characterize gapless states in Andreev interferometers, we need to compare transport processes in both arms of the circuit. We note that $\mathcal{W}_{heh}$, which is dominated by Andreev reflections and thus less sensitive to the presence of the SL-SR junction, shows negative correlations and a weak phase dependence~\cite{SM}. The behavior of all studied WTDs and correlations is summarized in \cref{tab:WTDs_All}. 

\textit{Concluding remarks.---} 
We have analyzed the distribution of waiting times and their correlations for electrons and holes emitted from a topological Andreev interferometer: a NL-SL-SR-NR junction on the quantum spin Hall edge. Two special features of this setup are (i) the emergence of gapless MBSs and (ii) it acting as an electron-hole beam splitter, sending holes and electrons to different leads. This topological Andreev interferometer is thus one of the simplest multi-terminal setups without magnetic elements that features MBSs and allows us to test their nonlocal behavior. 
We find that the gapless property of the MBSs makes the waiting times involving Majorana-assisted electron transfers, $\mathcal{W}_{ee}$ and $\mathcal{W}_{he}$, very special around $\phi\sim\pi$. Most importantly, the nonlocal property of MBSs is captured in the correlations between waiting times, see \cref{tab:WTDs_All}. For example, the waiting times for consecutive hole reflections and for alternate electron-hole detections are strongly correlated for gapless Majorana states [\cref{fig:jointWTD}(b), \cref{fig:jointWTD2}(b)], even if their distributions, $\mathcal{W}_{hh}$ and $\mathcal{W}_{eh}$, are almost insensitive to the phase $\phi$. 

\begingroup
\squeezetable
\begin{table}[t!]
	\begin{tabular}{|c|c|c|c|}
		\hline 
		WTD &  $\phi\!=\!0$ &  $\phi\!=\!\pi$ \\ \hline 
		$\mathcal{W}_{ee}$ & P  & Resonant level \\  \hline 
		$\mathcal{W}_{hh}$ & WD  & WD \\ \hline 
		$\mathcal{W}_{eh}$ & $\mathcal{W}_{eh}(0)\approx 0$, WD & $\mathcal{W}_{eh}(0)> 0$, WD  \\ \hline 
		$\mathcal{W}_{he}$ & $\mathcal{W}_{he}(0)\approx 0$, P & $\mathcal{W}_{he}(0)> 0$, P with dip \\ \hline  \hline
		$\delta\mathcal{W}_{eee}$ & Uncorrelated  & Uncorrelated \\  \hline 
		$\delta\mathcal{W}_{hhh}$ & Bream-splitter & $\delta\mathcal{W}_{hhh}\sim 1$ at $\tau_{1,2}<\mean{\tau_h}$ \\  \hline 
		$\delta\mathcal{W}_{heh}$ & Low correlation  & Low correlation \\  \hline 
		$\delta\mathcal{W}_{ehe}$ & Low correlation & $\delta\mathcal{W}_{ehe}\sim0.1$ at $\tau_{1}<\mean{\tau_h}$ \\  \hline 
	\end{tabular}
	\caption{Summary of the behaviors of WTDs and joint WTDs for $L_{\rm S}\!>\!\xi$. Here, WD and P indicate, respectively, Wigner-Dyson and Poisson distributions. }
	\label{tab:WTDs_All}
\end{table}
\endgroup

The search for signatures of Majorana states currently faces several challenges~\cite{Frolov2019,Prada2020,Frolov2021,Kleinherbers2023,Egues_2023}. In particular, the presence of trivial low-energy modes that mimic the local properties of MBSs, obscuring their detection in many platforms. 
To circumvent the latter, the quantum spin Hall effect is a promising platform where signatures of gapless MBSs have already been identified experimentally, even in the presence of extra trivial modes~\cite{Wiedenmann2016,Bocquillon2017,Deacon2017}. Thus, Andreev reflection and normal transmissions, which determine all our results, are the dominant scattering processes.
Furthermore, our analysis of nonlocal properties of Majorana states addresses the need to go beyond local signatures. In fact, previous works have already analyzed the WTD of electrons tunneling into a Majorana state~\cite{Chevallier2016,Mi2018,Fu2022,Schulz2022} focusing on its local properties and showing that the resonant transport through a MBS yields WTDs very similar to that of a single (trivial) level resonance~\cite{Haack2014,Mi2018}. 
Therefore, our results complement earlier work by showing that taking into account independent and simultaneous transfers of electrons and holes into separate detectors (i.e., both local and nonlocal WTDs), and the correlations between them, yields distinctive signatures of Majorana modes. Also, despite the challenges involved in measuring waiting times, there are promising new advances like experimental measurements of time-of-flight of electron excitations~\cite{Kataoka2016,Roussely2018}, in addition to recent theoretical proposals for WTD clocks~\cite{Haack2015,Dasenbrook2016}. 

\acknowledgments
Work at PRL is supported by Department of Space, Government of India. P.D.~thanks D.~Chakraborty, P.~Holmvall, and I.~Mahyaeh for technical help, U.~Basu for discussions, and acknowledges the computational facilities by Uppsala University and the Department of Science and Technology (DST), India, for the financial support through SERB Start-up Research Grant (File No.~SRG/2022/001121). 
J.C. acknowledges financial support from the Swedish Research Council (Vetenskapsr{\aa}det Grant No. 2021-04121), the Scandinavia-Japan Sasakawa Foundation (Grant No. GA22-SWE-0028), the Royal Swedish Academy of Sciences (Grant No. PH2022-0003), and the Carl Trygger’s Foundation (Grant No. 22: 2093). 
A.B.S. acknowledges financial support from the Swedish Research Council (Vetenskapsr{\aa}det Grant No. 2018-03488) and the Knut and Alice Wallenberg Foundation through the Wallenberg Academy Fellows program and the EU-COST Action CA-21144 Superqumap. 
P.B.~acknowledges support from the Spanish CM ``Talento Program'' project No.~2019-T1/IND-14088 and the Agencia Estatal de Investigaci\'on projects No.~PID2020-117992GA-I00 and No.~CNS2022-135950. 

\bibliography{bibfile}


\clearpage

\onecolumngrid
\setcounter{equation}{0}
\renewcommand{\theequation}{S\,\arabic{equation}}
\setcounter{figure}{0}
\renewcommand{\thefigure}{S\,\arabic{figure}}
%
%
\section{Supplementary material for ``Nonlocality of Majorana bound states revealed by electron waiting times in a topological Andreev interferometer"}

In this supplemental material, we provide details of the theoretical formalism and some additional results to support the discussions of the main text about the scattering probabilities, the bias dependence of the waiting time distributions, and the nonlocal joint waiting time correlations. 

\section{Theoretical methods} \label{sec:App-method}

In this section, we summarize the most important steps in order to arrive at the scattering matrices used in the main text and describe the theory for WTDs and joint WTDs.

\subsection{Scattering matrix formalism\label{apnd1}}

We consider a NL-SL-SR-NR junction, with NL and NR normal leads and SL and SR superconducting ones, along the x direction on one edge of a QSHI, where the NL-SL interface is placed at $x=0$, the SL-SR interface at $x=L_\text{S}$, and the SR-NR one at $x=2L_\text{S}$. The scattering states at the different regions of the NL-SL-SR-NR junction can be written as~\cite{Cayao2022}
\begin{subequations}\label{phiNS}
\begin{align}
\psi_{\rm 1}(x)={}& \left\{ \begin{array}{cl}
 \phi_{1}^{N}\,{\rm e}^{ik_{e}x}+r_{\rm eh}^{\rm L}\phi_{3}^{N}\,{\rm e}^{ik_{h}x} +r_{\rm ee}^{\rm L}\phi_{2}^{N}\,{\rm e}^{-ik_{e}x}, & x<0  \\
  a_{1}\phi_{1}^{\rm S}\,{\rm e}^{ik^{\rm S}_{e}x}+b_{1}\phi_{2}^{\rm S}\,{\rm  e}^{-ik^{\rm S}_{e}x}+c_{1}\phi_{3}^{\rm S}\,{\rm e}^{ik^{\rm S}_{h}x}+d_{1}\phi_{4}^{\rm S}\,{\rm e}^{-ik^{\rm S}_{h}x} , & 0<x<L_{\rm S} \\
  p_{1}\phi_{1}^{\rm S}\,{\rm e}^{ik^{\rm S}_{e}x}+q_{1}\phi_{2}^{\rm S}\,{\rm e}^{-ik^{\rm S}_{e}x}+r_{1}\phi_{3}^{\rm S}\,{\rm e}^{ik^{\rm S}_{h}x}+s_{1}\phi_{4}^{\rm S}\,{\rm e}^{-ik^{\rm S}_{h}x} , & L_{\rm S}<x<2L_{\rm S} \\   
 t_{\rm ee}^R  \phi_{1}^{\rm N}\,{\rm e}^{ik_{e}x} + t_{\rm eh}^{\rm R}  \phi_{4}^{\rm N}\,{\rm e}^{-ik_{h}x}, & x>2L_{\rm S} 
 \end{array} \right. ,
\\ 
\psi_{2}(x)={}& \left\{ \begin{array}{cl}
\phi_{4}^{N}\,{\rm e}^{-ik_{h}x}++r_{\rm he}^{\rm L}\phi_{2}^{N}\,{\rm e}^{-ik_{e}x} +r_{\rm hh}^{\rm L} \phi_{3}^{N}\,{\rm e}^{ik_{h}x}, & x<0 \\
a_{2}\phi_{1}^{\rm S}\,{\rm e}^{ik^{\rm S}_{e}x}+ b_{2}\phi_{2}^{\rm S}\,{\rm e}^{-ik^{\rm S}_{e}x}+ c_{2}\phi_{3}^{\rm S}\,{\rm e}^{ik^{\rm S}_{h}x}+ d_{2}\phi_{4}^{\rm S}\,{\rm e}^{-ik^{\rm S}_{h}x}, & 0<x<L_{\rm S}  \\
p_{2}\phi_{1}^{\rm S}\,{\rm e}^{ik^{\rm S}_{e}x}+q_{2}\phi_{2}^{\rm S}\,{\rm e}^{-ik^{\rm S}_{e}x}+r_{2}\phi_{3}^{\rm S}\,{\rm e}^{ik^{\rm S}_{h}x}+s_{2}\phi_{4}^{\rm S}\,{\rm e}^{-ik^{\rm S}_{h}x}, & L_{\rm S}<x<2L_{\rm S} \\ 
t_{\rm hh}^{\rm R}  \phi_{4}^{\rm N}\,{\rm e}^{-ik_{h}x} + t_{\rm he}^{\rm R}  \phi_{1}^{\rm N}\,{\rm e}^{ik_{e}x}, \,x>2L_{\rm S} 
 \end{array} \right. ,
\\
\psi_{3}(x)={}& \left\{ \begin{array}{cl}
c_{3}  \phi_{2}^{N}\,{\rm e}^{-ik_{e}x} +d_{3}  \phi_{3}^{N}\,{\rm e}^{ik_{h}x}, &  x<0  \\
a_{3}\phi_{1}^{\rm S}\,{\rm e}^{ik^{\rm S}_{e}x}+b_{3}\phi_{2}^{\rm S}\,{\rm e}^{-ik^{\rm S}_{e}x}+c_{3}\phi_{3}^{\rm S}\,{\rm e}^{ik^{\rm S}_{h}x}+d_{3}\phi_{4}^{\rm S}\,{\rm e}^{-ik^{\rm S}_{h}x}, & 0<x<L_{\rm S} \\
p_{3}\phi_{1}^{\rm S}\,{\rm e}^{ik^{\rm S}_{e}x}+q_{3}\phi_{2}^{\rm S}\,{\rm e}^{-ik^{\rm S}_{e}x}+r_{3}\phi_{3}^{\rm S}\,{\rm e}^{ik^{\rm S}_{h}x}+s_{3}\phi_{4}^{\rm S}\,{\rm e}^{-ik^{\rm S}_{h}x}, & L_{\rm S}<x<2L_{\rm S} \\ 
\phi_{2}^{\rm N}\,{\rm e}^{-ik_{e}x}+a_{3}\phi_{4}^{\rm N}\,{\rm e}^{-ik_{h}x}+b_{3}\phi_{1}^{\rm N}\,{\rm e}^{ik_{e}x}, & x>2L_{\rm S} 
 \end{array} \right. ,
\\
\psi_{4}(x)={}& \left\{ \begin{array}{cl}
c_{4}  \phi_{3}^{N}\,{\rm e}^{ik_{h}x} + d_{4}  \phi_{2}^{N}\,{\rm e}^{-ik_{e}x}, & x<0   \\
a_{4}\phi_{1}^{\rm S}\,{\rm e}^{ik^{\rm S}_{e}x}+ b_{4}\phi_{2}^{\rm S}\,{\rm e}^{-ik^{\rm S}_{e}x}+ c_{4}\phi_{3}^{\rm S}\,{\rm e}^{ik^{\rm S}_{h}x}+ d_{4}\phi_{4}^{\rm S}\,{\rm e}^{-ik^{\rm S}_{h}x}, & 0<x<L_{\rm S} \\
p_{4}\phi_{1}^{\rm S}\,{\rm e}^{ik^{\rm S}_{e}x}+ q_{4}\phi_{2}^{\rm S}\,{\rm e}^{-ik^{\rm S}_{e}x}+r_{4}\phi_{3}^{\rm S}\,{\rm e}^{ik^{\rm S}_{h}x}+s_{4}\phi_{4}^{\rm S}\,{\rm e}^{-ik^{\rm S}_{h}x}, & L_{\rm S}<x<2L_{\rm S} \\ 
\phi_{3}^{\rm N}\,{\rm e}^{ik_{h}x}+a_{4}\phi_{1}^{\rm N}\,{\rm e}^{ik_{e}x}+b_{4}\phi_{4}^{\rm N}\,{\rm e}^{-ik_{h}x}, & x>2L_{\rm S} 
 \end{array} \right. ,
\end{align}
\end{subequations}
where
\beq
\begin{split}
\phi_{1}^{\rm N}\!&=\!\begin{pmatrix}
1,0,0,0
 \end{pmatrix}^{T},
\phi_{2}^{\rm N}\!=\!\begin{pmatrix}
0,1,0,0
 \end{pmatrix}^{T},
\phi_{3}^{\rm N}\!=\!\begin{pmatrix}
0,0,1,0
 \end{pmatrix}^{T},
\phi_{4}^{N}\!=\!\begin{pmatrix}
0,0,0,1
 \end{pmatrix}^{T},\\
 \phi_{1}^{\rm S}\!&=\!\begin{pmatrix}
u, 0,v,0
 \end{pmatrix}^{T},
\phi_{2}^{\rm S}\!=\!\begin{pmatrix}
0,u,0,v
 \end{pmatrix}^{T},
\phi_{3}^{\rm S}\!=\!\begin{pmatrix}
v,0,u,0
 \end{pmatrix}^{T},
\phi_{4}^{\rm S}\!=\!\begin{pmatrix}
0,v,0,u
 \end{pmatrix}^{T}.
\end{split} \non
\eeq
The $a_i$, $b_i$, $c_i$, and $d_i$ are the scattering parameters to be determined. The wave vectors in normal regions NL and NR read as 
\bea
k_{e(h)}(E,\Delta)=\frac{\mu \pm E}{v_F},
\eea
and in the superconducting regions SL and SR take the form,
\bea
k_{e(h)}^S(E,\Delta)=\frac{\mu}{v_F}\pm\frac{\sqrt{E^2-\Delta^2}}{v_F} .
\eea
We solve the above equations by matching the wave functions at the interfaces $x=0$, $x=L_{\rm S}$, and $x=2L_{\rm S}$, to obtain 
\bea
b_{i}&=0\,,\quad
d_{i}=0\,,\quad
p_{2,3}=0\,,\quad
q_{1,4}=0\,,\quad
r_{2,3}=0\,, \quad
s_{1,4}&=0\,.
\eea
The local normal reflection and nonlocal crossed Andreev transmission are forbidden by the conservation of helicity of the edge states of the QSHI~\cite{Adroguer2010,Tkachov2013} allowing only two processes: (i) local Andreev reflections, where an incident electron is reflected as a hole at NL, and (ii) nonlocal electron transmission at NR with the amplitudes as follows~\cite{Cayao2022}. 
\noindent
\begin{itemize}
\item Andreev reflection in NL for an incident electron from NL:
\beq\label{eq:ref-amp}
r^{\rm L}_{\rm eh}=\frac{(e^{i k_{\rm e}^{\rm S} L_{\rm S}} - e^{i k_{\rm h}^{\rm S} L_{\rm S}}) uv [(e^{i k_{\rm e}^{\rm S} L_{\rm S}} + e^{i (k_{\rm h}^{\rm S} L_{\rm S} + \phi)}) u^2 - (e^{i k_{\rm h}^{\rm S} L_{\rm S}} +e^{i (k_{\rm e}^{\rm S} L_{\rm S} + \phi)})v^2]}{ (e^{i k_{\rm e}^{\rm S} L_{\rm S}}-e^{i k_{\rm h}^{\rm S} L_{\rm S}})^2 u^2 v^2 - (e^{ik_{\rm h}^{\rm S} L_{\rm S}} u^2 -e^{i  k_{\rm e}^{\rm S} L_{\rm S}} v^2)^2 \,e^{i \phi}}.
\eeq
\item Transmission of an electron into NR for an incident electron from NL:
\beq\label{eq:tran-amp}
t_{\rm ee}^{\rm R}=-\frac{(u^2-v^2)^2 e^{-2i(k_{\rm e}-k_{\rm e}^{\rm S}-k_{\rm h}^{\rm S}) L_{\rm S}}e^{i\phi}} {[e^{ik_{\rm e}^{\rm S} L_{\rm S}}-e^{ik_{\rm h}^{\rm S}L_{\rm S}}]^2 u^2 v^2 - [u^2 e^{i k_{\rm h}^{\rm S}L_{\rm S}}-v^2 e^{i k_{\rm e}^{\rm S} L_{\rm S}}]^2 e^{i\phi}}.
\eeq
\end{itemize}
The amplitudes for the hole counterpart can be found by taking the complex conjugate of the amplitudes for the electron part with inversion of energy. Finally we construct the $\mathcal{S}$-matrix as
\beq
\mathcal{S} =
\begin{pmatrix}
r^{\rm L}_{\rm ee} &r^{\rm L}_{\rm eh}& t^{\prime \rm R}_{\rm ee}& t^{\prime \rm R}_{\rm eh} \\
r^{\rm L}_{\rm he} &r^{\rm L}_{\rm hh}& t^{\prime \rm R}_{he} & t^{\prime \rm R}_{\rm hh} \\
t_{\rm ee}^{\rm R} & t_{\rm eh}^{\rm R}& r_{\rm ee}^{\prime \rm L}& r_{\rm hh}^{\prime \rm L} \\
t_{\rm he}^{\rm R}  & t_{\rm hh}^{\rm R} & r_{\rm he}^{\prime \rm L}& r_{\rm hh}^{\prime \rm L}
\end{pmatrix} .
\label{smatwhole}
\eeq 
Now, due to the helicity of the edge states, we have $r^{\rm L}_{\rm ee}\!=\!r^{\rm L}_{\rm hh}\!=\!r^{\prime \rm L}_{\rm ee}\!=\!r^{\prime \rm L}_{\rm hh}\!=\!t^{\rm R}_{\rm eh}\!=\!t^{\rm L}_{\rm he}\!=\!t^{\prime \rm R}_{\rm eh}\!=\!t^{\prime \rm R}_{\rm he}\!=\!0$. After rotating the basis, we arrive at the block-diagonal form of the $\mathcal{S}$-matrix which we use to evaluate the WTDs. We also rename the components of the $\mathcal{S}$-matrix here, which we have used in the main text, 
\beq
\mathcal{S} =\begin{pmatrix} t^{\prime\,R}_{ee} &r^L_{eh}&0 & 0 \\
 r_{he}^{\prime\,L}  & t^R_{hh}&  0 & 0 \\
 0& 0 & t^R_{ee}& r_{eh}^{\prime\,L} \\
0& 0 & r^L_{he} & t^{\prime\,R}_{hh}
\end{pmatrix}=\begin{pmatrix} \mathcal{S}^{\prime}_{eL,eR} &\mathcal{S}_{eL,hL}&0 & 0 \\
 \mathcal{S}_{hL,eL}^{\prime}  & \mathcal{S}_{hR,hR}&  0 & 0 \\
 0& 0 & \mathcal{S}_{eL,eR}& \mathcal{S}_{eL,hL}^{\prime} \\
0& 0 & \mathcal{S}_{hL,eL} & \mathcal{S}^{\prime}_{hL,hR}
\end{pmatrix}.
\label{smat}
\eeq 

\subsection{Waiting time distributions\label{apnd2}}

The waiting time is a fluctuating quantity, which must be described by a probability distribution. The waiting time distribution (WTD) is the conditional probability density of detecting a particle of type $\beta$ at time $t_\beta^e$, given that the last detection of a particle of type $\alpha$ occurred at the earlier time $t_\alpha^s$. Here, the types $\alpha$ and $\beta$ may refer to the out-going channel, the spin of the particle, and the particle being an electron or a hole. The WTD is denoted as $\mathcal{W}_{\alpha\rightarrow\beta}(t_\alpha^s,t_\beta^e)$. For the systems considered here, with no explicit time dependence, the WTD is a function only of the time difference, such that $\mathcal{W}_{\alpha\rightarrow\beta}(t_\alpha^s,t_\beta^e)= \mathcal{W}_{\alpha\rightarrow\beta}(\tau)$ with $\tau=t_\beta^e-t_\alpha^s$. 


To evaluate the WTD, we proceed as in Ref.~\cite{Dasenbrook2015} and express the WTD as time-derivatives of the idle time probability. The idle time probability $\Pi(t^s_\alpha,t^{e}_\alpha)$ is the probability that no particles of type $\alpha$ are detected in the time interval $[t^s_\alpha,t^{e}_\alpha]$ by a detector at position $x_\alpha$. The idle time probability can be a function of several different particle types and associated time intervals. 

The idle time probability can be evaluated using scattering theory, leading to the determinant formula~\cite{Dasenbrook2015}
\begin{equation}
	\Pi(\{t^{s}_\gamma, t^{e}_\gamma\})= \det[\mathbbm{1} - \mathcal{Q}(\{t^{s}_{\gamma}, t^{e}_{\gamma} \})] ,
	\label{eq:ITP}
\end{equation}
where the set $\{t^{s}_\gamma, t^{e}_\gamma\}$ corresponds to all relevant particles $\gamma$ and associated time intervals. The Hermitian operator $\mathcal{Q}(\{t^{s}_\gamma, t^{e}_\gamma\})$ is a matrix in a combined energy and particle type representation. It has the block form
\begin{equation}
	[\mathcal{Q}]_{EE'} = \mathcal{S}^{\dagger}(E)\, \mathcal{K}(E-E')\,\mathcal{S}(E'),
	\label{eq:qmatrix}
\end{equation}
having omitted the time arguments. The scattering matrix $\mathcal{S}(E)$, for particles with excitation energy $E$, and the kernel $\mathcal{K}(E)$ are matrices in the space of particle types. The kernel is the block diagonal matrix
\begin{equation}
	\mathcal{K}(\{t^{s}_\gamma, t^{e}_\gamma\});E) =\bigoplus_{\gamma} K( t^{s}_\gamma, t^{e}_\gamma; E)\label{eq:K-kernels} ,
\end{equation}
given by the direct sum of kernels
\begin{equation}
	K(t^{s}_\gamma, t^{e}_\gamma; E)
	=\frac{\kappa}{\pi E} e^{-i \frac{E}{2} \left(t^{e}_\gamma+t^{s}_\gamma+\frac{2x_\gamma}{\hbar v_F}\right)}
	\sin \left[ \frac{E}{2}(t^{e}_\gamma-t^s_\gamma) \right],
	\label{eq:kernel-app}
\end{equation}
corresponding to each particle of type $\gamma$ detected at position $x_\gamma$. \Cref{eq:kernel-app} corresponds to Eq.~(3) in the main text. 
We work close to the Fermi level, where the dispersion relation $E\!=\!\hbar v_Fk$ is linear and all quasi-particles propagate with the Fermi velocity~$v_F$. To implement the matrix in \cref{eq:qmatrix}, we discretize the transport window $[E_F,E_F+eV]$ in $N$ intervals of width $\kappa\!=\!eV/N$. The width $\kappa$ explicitly enters in \cref{eq:kernel-app}, and we always consider the limit $N\!\rightarrow\!\infty$, for which the transport is stationary. 

The WTD can be related to the idle time probability by realizing that time derivatives correspond to detection events~\cite{Haack2014}. When taking derivatives of operators and determinants, we use Jacobi's formula
\begin{equation}
	\frac{\mathrm{d}}{\mathrm{d}\tau} \det\{\check{A}(\tau)\} = \det\{\check{A}(\tau)\} \mathrm{Tr} \{\check{A}^{-1}(\tau) \frac{\mathrm{d}}{\mathrm{d}\tau} \check{A}(\tau)\} .
	\label{eq:Jacobi}
\end{equation}
Applied to \cref{eq:ITP}, the derivative takes the form
\begin{equation}
	\partial_{t_\alpha} \Pi(\{t^{s}_\gamma, t^{e}_\gamma\}) = - \Pi(\{t^{s}_\gamma, t^{e}_\gamma\}) \mathrm{Tr} \{ \mathcal{G}(\{t^{s}_\gamma, t^{e}_\gamma\})  \dot{\mathcal{Q}}_\alpha (t_\alpha) \} ,
	\label{eq:deriv}
\end{equation}
where we have defined 
\begin{equation}
	\mathcal{G}(\{t^{s}_\gamma, t^{e}_\gamma\})= [\mathbbm{1} - \mathcal{Q}(\{t^{s}_\gamma, t^{e}_\gamma\}) ]^{-1}
\end{equation}
and $ \dot{\mathcal{Q}}_\alpha = \partial_{t_\alpha} \mathcal{Q} $. After recasting the kernel definition, Eq.\,[3] of the maintext and \cref{eq:kernel-app}, as 
\begin{equation}
	K(t^{s}_\gamma, t^{e}_\gamma; E)
	=\frac{\kappa}{2 i \pi E} e^{-i \frac{E x_\gamma}{\hbar v_F} }
	\left( e^{-i E t^{e}_\gamma/\hbar } - e^{-i E t^{s}_\gamma / \hbar } \right),
	\label{eq:kernel2}
\end{equation}
it is straightforward to see that the derivative with respect to time $t_\alpha$ of the operator $\mathcal{Q}(\{t^{s}_\gamma, t^{e}_\gamma\}) $ depends only on time $t_\alpha$, that is, $ \dot{\mathcal{Q}}_\alpha = \dot{\mathcal{Q}}_\alpha (t_\alpha) $. Consequently, we have 
\begin{equation}
	\ddot{\mathcal{Q}}_{\alpha\beta} = \frac{\partial^2}{\partial_{t_\alpha} \partial_{t_\beta}}\mathcal{Q}(\{t^{s}_\gamma, t^{e}_\gamma\}) = \delta_{\alpha\beta} \frac{\partial^2}{\partial_{t_\beta}^2}\mathcal{Q}(\{t^{s}_\gamma, t^{e}_\gamma\}) . 
	\label{eq:dob-der}
\end{equation}

The first-passage time distributions are defined as 
\begin{equation}
	\tilde{\mathcal{F}}_{\alpha}(\tau) = \partial_{t^{s}_{\alpha}} \Pi\big|_{t^{s}_\alpha,t^s_\beta,t^e_\beta,\rightarrow 0}^{t^{e}_\beta\rightarrow \tau} = -\Pi\, \left. \mathrm{Tr}\left\{ \mathcal{G}\frac{\partial \mathcal{Q}}{\partial t^{s}_{\alpha}}\right\} \right|_{t^{s}_\alpha,t^s_\beta\rightarrow 0}^{t^e_\alpha\rightarrow 0, t^{e}_\beta\rightarrow \tau} 
	\label{eq:FPT1}
\end{equation}
and
\begin{equation}
	\mathcal{F}_{\beta}(\tau) =-\partial_{t^{e}_{\beta}}\Pi\big|_{t^{s}_\alpha,t^s_\beta,t^e_\beta,\rightarrow 0}^{t^{e}_\beta\rightarrow \tau}=\Pi\,  \left. \mathrm{Tr}\left\{\mathcal{G} \frac{\partial \mathcal{Q}}{\partial t^{e}_{\beta}}\right\} \right|_{t^{s}_\alpha,t^s_\beta\rightarrow 0}^{t^e_\alpha\rightarrow 0, t^{e}_\beta\rightarrow \tau} .
	\label{eq:FPT2}
\end{equation}
\Cref{eq:FPT1} is the conditional probability density that no particles of type $\beta$ are detected in the time span $[0,\tau]$, given that a particle of type $\alpha$ was detected at the initial time $t\!=\!0$. Similarly, \cref{eq:FPT2} concerns the time $\tau$ we have to wait until a particle of type $\alpha$ is detected, given that we start the clock at time $t\!=\!0$. 
For the single-channel stationary case, where all quantities depend only on the time difference $\tau= t^e-t^s$ and we have only one particle type, we find that these two probability distributions are the same, 
\[
\tilde{\mathcal{F}}_\alpha(\tau) = \mathcal{F}_\alpha(\tau) = -\partial_\tau \Pi(\tau) .
\]

The distribution of waiting times between particles of type $\alpha$ and particles of type $\beta$ can be expressed as\,\cite{Dasenbrook2015}
\begin{equation}
	I_{\alpha} \mathcal{W}_{\alpha \rightarrow \beta}(\tau)  = -\partial_{t^{e}_\beta} \partial_{t^{s}_\alpha}\Pi(t_{\alpha}^s,t_{\alpha}^e;t_{\beta}^s,t_\beta^e\})\big|_{t^{s}_\alpha,t^s_\beta \rightarrow 0}^{t^e_\alpha \rightarrow 0, t^{e}_\beta\rightarrow \tau} \,,
	\label{eq:WTD}
\end{equation}
where $I_\alpha$ is the average particle current of type $\alpha$ particles, and the minus sign comes together with the derivative with respect to the starting time $t_\alpha^s$. In addition, after having performed the derivatives, we set the starting times to zero, i.e., $t_\alpha^s\!=\!t_\beta^s=0$, while for the end times we set $t^{e}_\alpha\!=\! 0$ and $t^{e}_\beta\!=\!\tau$.  The waiting time is then measured from the time when a particle of type $\alpha$ is detected until the later time when a particle of type $\beta$ is detected. During this waiting time, additional particles of type $\alpha$ may be detected, but not of type $\beta$.

By combining \cref{eq:WTD,eq:ITP}, we find
\begin{equation}
	I_{\alpha} \mathcal{W}_{{\alpha} \rightarrow {\beta}}(\tau)=\frac{1}{\Pi}\tilde{\mathcal{F}}_{\alpha}\mathcal{F}_{\beta}
	+ \left. \Pi \, \mathrm{Tr}\left\{ \mathcal{G} \frac{\partial \mathcal{Q}}{\partial t^{e}_{\beta}}\mathcal{G} \frac{\partial \mathcal{Q}}{\partial t^{s}_{\alpha}} \right\} \right|_{t^{s}_\alpha,t^s_\beta\rightarrow 0}^{t^e_\alpha\rightarrow 0, t^{e}_\beta\rightarrow \tau} \,,
	\label{eq:WTDfinal}
\end{equation}
having made repeatedly use of Jacobi's formula for derivatives of determinants. Finally, for evaluating \cref{eq:WTDfinal} we note that the average particle current of type $\alpha$ particles can be expressed as $I_\alpha\!=\!\mathcal{F}_{\alpha}(0)$. In combination, Eqs.~(\ref{eq:ITP}-\ref{eq:WTDfinal}) allow us to evaluate the distributions of waiting times for the superconducting systems that we consider in the main text. We have plotted them as a function of $\tau_i$ normalized by the mean time $\langle \tau_i \rangle$. 

\subsection{Joint waiting time distributions}

We can now introduce the joint waiting time distribution $\mathcal{W}_{\alpha \rightarrow \gamma \rightarrow \beta}(t^s_\alpha,t^m_\gamma,t^e_\beta)$. This probability distribution generalizes the waiting time distribution between particles of type $\alpha$ and $\beta$, \cref{eq:WTD,eq:WTDfinal}, to include the extra detection of a particle of type $\gamma$ at an intermediate time $t^m_\gamma$, such that $t^s_\alpha<t^m_\gamma<t^e_\beta$. 
To find the joint waiting time distribution, we introduce an auxiliary, \textit{virtual} particle channel of type $\gamma$. This new channel allows us to perform a third derivative, representing the intermediate detection event, by adding an extra kernel $K(t^s_\gamma,t^e_\gamma)$ to \cref{eq:ITP,eq:K-kernels}. Eventually, we set $t^s_\gamma=t^e_\gamma=t^m$, effectively closing this channel's contribution to the idle time probability. 

For a system with two particle types, with idle time probability $\Pi(t^s_1,t^e_1;t^s_2,t^e_2)$, the joint waiting time distribution can be defined as~\cite{Dasenbrook2015}
\begin{equation}
	I_{\alpha} \mathcal{W}_{\alpha \rightarrow \gamma \rightarrow \beta}(t^s,t^m,t^e)  = \partial_{t^{e}_\beta} \partial_{t^{e}_\gamma} \partial_{t^{s}_\alpha} \Pi( t_{\alpha}^s,t_{\alpha}^e;t_{\beta}^s,t_\beta^e;t_{\gamma}^s,t_{\gamma}^e ) 
	\big|_{t^{s}_\alpha \rightarrow t^s,t^{s}_\gamma \rightarrow t^m,t^s_\beta \rightarrow t^s}^{t^e_\alpha \rightarrow t^s, t^e_\gamma \rightarrow t^m, t^{e}_\beta\rightarrow t^e}  ,
	\label{eq:JWTD}
\end{equation}
where we have extended the idle time probability to include the auxiliary channel. From \cref{eq:dob-der}, second partial derivatives over the auxiliary channel are zero. In fact, all second derivatives are zero, since $t^s_\alpha$ and $t^e_\beta$ are independent variables even when $\alpha=\beta$. Consequently, the joint waiting time distribution has the following structure:
\begin{align}
	\partial_3 \partial_2 \partial_1 \Pi = {}& 
	-\Pi \mathrm{Tr}\{ \mathcal{G} \dot{Q}_3 \} \mathrm{Tr}\{ \mathcal{G} \dot{Q}_2 \} \mathrm{Tr}\{ \mathcal{G} \dot{Q}_1 \} 
	- \Pi \mathrm{Tr}\{ \mathcal{G} \dot{Q}_3 \mathcal{G} \dot{Q}_2 \mathcal{G} \dot{Q}_1 + \mathcal{G} \dot{Q}_3 \mathcal{G} \dot{Q}_1 \mathcal{G} \dot{Q}_2  \} \notag \\
	& + \Pi \mathrm{Tr}\{ \mathcal{G} \dot{Q}_3 \} \mathrm{Tr}\{ \mathcal{G} \dot{Q}_2 \mathcal{G} \dot{Q}_1 \} 
	+ \Pi \mathrm{Tr}\{ \mathcal{G} \dot{Q}_2 \} \mathrm{Tr}\{ \mathcal{G} \dot{Q}_3 \mathcal{G} \dot{Q}_1 \}
	+ \Pi \mathrm{Tr}\{ \mathcal{G} \dot{Q}_1 \} \mathrm{Tr}\{ \mathcal{G} \dot{Q}_3 \mathcal{G} \dot{Q}_2 \} ,
	\label{eq:jointWTD-gen}
\end{align}
where $\partial_{j}$, with $j=1,2,3$, are time derivatives over some generic and independent variables. It is important to stress that both $\Pi$ and $\mathcal{G}$ in \cref{eq:jointWTD-gen} depend only on the time coordinates of the original channels, since both $t^{s,e}_\gamma\rightarrow t^m$. That is, $\Pi=\Pi(t_{\alpha}^s,t_{\alpha}^e;t_{\beta}^s,t_\beta^e)$ and $\mathcal{G}=\mathcal{G}(t_{\alpha}^s,t_{\alpha}^e;t_{\beta}^s,t_\beta^e)$. The only dependence on $t^m$ is thus on the intermediate derivative, $\dot{Q}_2$. 

Based on the general structure of \cref{eq:jointWTD-gen}, we denote as $\gamma$ the auxiliary channel where the intermediate detection event takes place, and define the same channel joint waiting time distribution
\begin{equation}
	I_{\alpha} \mathcal{W}_{\alpha \rightarrow \gamma \rightarrow \alpha}(\tau_1,\tau_2)  = \partial_{t^{e}_\alpha} \partial_{t^{e}_\gamma} \partial_{t^{s}_\alpha} \Pi(t_{\alpha}^s,t_{\alpha}^e;t_{\beta}^s,t_\beta^e;t_{\gamma}^s,t_{\gamma}^e) 
	\big|_{t^{s}_\alpha \rightarrow 0, t^s_\gamma \rightarrow \tau_1, t^{s}_\beta \rightarrow 0}^{t^e_\alpha \rightarrow \tau_1+\tau_2, t^{e}_\gamma \rightarrow \tau_1, t^{e}_\beta \rightarrow 0}  ,
	\label{eq:JWTD11}
\end{equation}
and the different channel joint waiting time distribution
\begin{equation}
	I_{\alpha} \mathcal{W}_{\alpha \rightarrow \gamma \rightarrow \beta}(\tau_1,\tau_2)  = \partial_{t^{e}_{\beta}} \partial_{t^{e}_\gamma} \partial_{t^{s}_\alpha} \Pi(t_{\alpha}^s,t_{\alpha}^e;t_{\beta}^s,t_\beta^e;t_{\gamma}^s,t_{\gamma}^e ) 
	\big|_{t^{s}_\alpha \rightarrow 0, t^s_\gamma \rightarrow \tau_1, t^{s}_\beta \rightarrow 0}^{t^e_\alpha \rightarrow 0, t^{e}_\gamma \rightarrow \tau_1, t^{e}_\beta \rightarrow \tau_1+\tau_2}  ,
	\label{eq:JWTD12}
\end{equation}
with $\beta\neq\alpha$. 
Here, we have assumed a stationary case and set the initial time $t^s_\alpha=0$. The intermediate detection time interval is then $t^m-t^s_\alpha=\tau_1$, and the total time interval is $t^e_{\alpha,\beta}-t^s_\alpha=\tau_1+\tau_2$. 


\section{\label{sec:apnd3}Scattering probabilities} 

In this section, we show some results of the scattering probabilities in our interferometer to support our discussions on WTDs in the main text.
\begin{figure}[!thpb]
\centering
\includegraphics[scale=0.6]{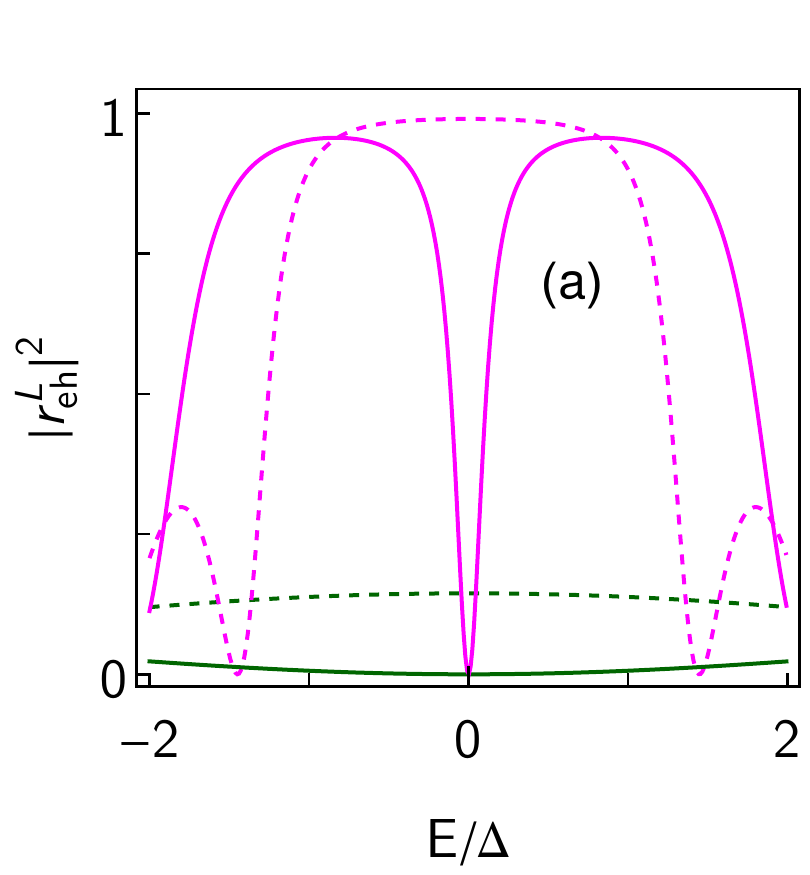}~~~
\includegraphics[scale=0.6]{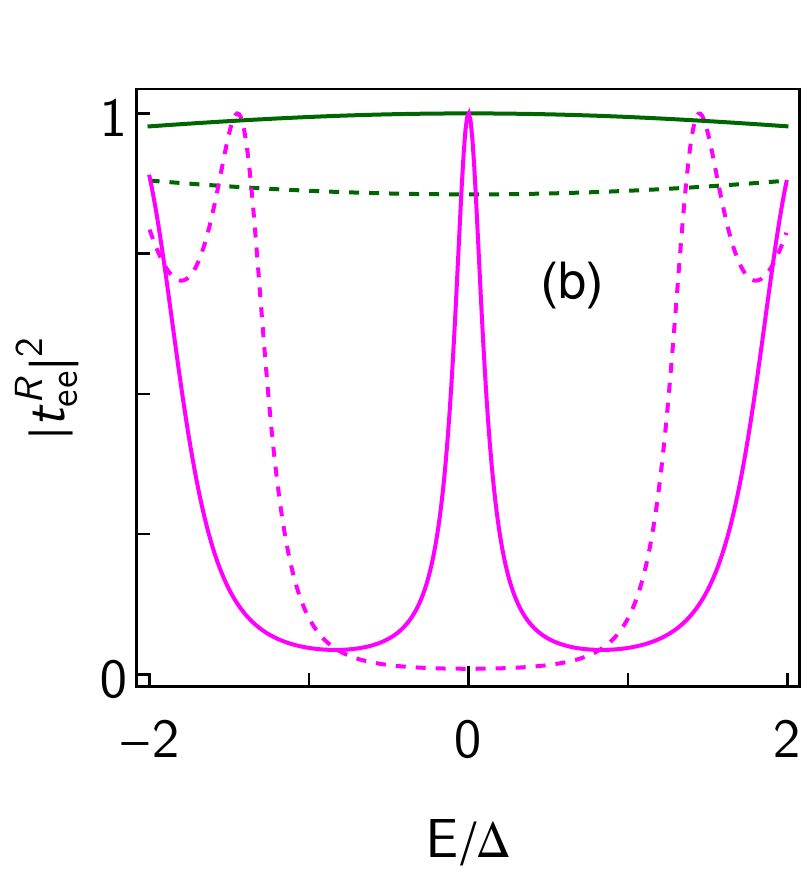}
\caption{(a) Reflection ($|r_{\rm eh}^{\rm L}|^2$) and (b) transmission ($|t_{\rm ee}^{\rm R}|^2$) probabilities as a function of the energy for $L_{\rm S}=0.2$ (green) and $1.5$ (magenta) at $\phi\!=\!0$ (dashed lines) and $\phi\!=\!\pi$ (solid lines).}
\label{fig:amplitude}
\end{figure}

In \cref{fig:amplitude}, we plot the energy dependence of the Andreev reflection probability $|r^{\rm L}_{\rm eh}|^2$, see \cref{eq:ref-amp}, at the left lead NL and the transmission $|t^{\rm R}_{\rm ee}|^2$, see \cref{eq:tran-amp}, at the right lead NR for $\phi\!=\!0$ and $\phi\!=\!\pi$. 
At $\phi\!=\!0$, we see that the Andreev reflection amplitude is constant (close to unity) within the superconducting gap $\Delta$ and decreases outside of the gap, followed by oscillations. The Andreev reflection for a long junction $L_{\rm s}\!\ge\!\xi$, \cref{fig:amplitude}(a), is similar to that of an ordinary normal metal-superconductor junction~\cite{Blonder1982}. Within the superconducting gap, Andreev reflection is the most likely process and thus its probability is close to unity. However, the Andreev reflection amplitude in our QSHI-based Andreev interferometer is very sensitive to the lengths of the two superconductors, $L_\text{S}$. As soon as we set $L_{\rm S}<\xi$, the probability of Andreev reflection decreases. The more we shorten the superconductor, the more the probability deviates from unity. For $L_{\rm S}\!\ll\!\xi$, it is highly flattened even beyond the gap. The mostly constant energy dependence of the probabilities for $\phi=0$ explains why their related WTDs, shown with dashed lines in Fig.~(2) of the main text, follow with little deviation the characteristic WTDs of energy-independent Poisson or Wigner-Dyson distributions. 

Interestingly, in addition to its dependence with $L_{\rm S}$, the Andreev reflection probability is also sensitive to the phase difference between the two superconductors. In the presence of a finite phase difference, the probability deviates strongly from unity around $E\!=\!0$, dropping to zero at $\phi\!=\!\pi$. This phenomenon is true for all values of $L_{\rm S}$, but the dip is particularly sharp for long junctions with $L_\text{S}>\xi$. This is a result of the formation of MBS at the junction, protected by the topology of the system. The phase tunable property of the MBS at the QSHI-based normal-superconductor junction is reflected in these profiles as presented in \cref{fig:amplitude}(a). 

Due to the electron/hole separation at the edges of QSHIs, the transmission spectra provide a picture complementary to the Andreev reflection. The transmission amplitude of an electron to the right lead NR, \cref{fig:amplitude}(b), is very low throughout the entire superconductor gap region, then increases for $E$ outside the gap at $\phi\!=\!0$. Again, this is similar to an ordinary normal metal-superconductor junction~\cite{Blonder1982}. The transmission probability increases when with the lengths of the two superconductors. It becomes more dramatic when we tune the phase to $\phi=\pi$. Around $E\!=\!0$, the transmission curves rise to unity with the sharpness of the peaks controlled by $L_{\rm S}$. 

The length and phase sensitivity of Andreev reflection and electron transmission amplitudes are nicely captured by the WTDs, as discussed in the main text. Even though both normal $|t^{\rm R}_{\rm ee}|^2$ and Andreev $|r^{\rm L}_{\rm eh}|^2$ probabilities feature a strong energy dependence for $\phi=\pi$, only the distribution between electron detections, $\mathcal{W}_{ee}$, shows a behavior different than the energy-independent WTDs, cf. Fig.~(2) of the main text. In particular, $\mathcal{W}_{ee}$ approximates the distribution for a single resonance~\cite{Mi2018}. We explain this difference with $\mathcal{W}_{hh}$ as follows: for sufficiently long junctions, $L_\text{S}>\xi$, $|t^{\rm R}_{\rm ee}|^2$ presents a narrow resonance at $E\sim0$ and $|r^{\rm L}_{\rm eh}|^2$ an equally narrow dip. However, any voltage bias bigger than the resonance width ($eV=\Delta/2$ in the main text) always features almost unitary transmission for some energies close to zero and almost zero reflection probability for the same energies, and, for energies away from zero but still within the gap, $|t^{\rm R}_{\rm ee}|^2$ and $|r^{\rm L}_{\rm eh}|^2$ quickly evolve to zero and one, respectively. Consequently, as long as the voltage bias is not fine-tuned to the edge of the resonance, the WTD for electrons is more sensitive to the energy dependence since it features only a few energy intervals with high probability for $E\sim0$ and the rest are almost zero. By contrast, $\mathcal{W}_{hh}$ follows the opposite behavior with relatively more energy intervals with high probability than those with zero.

\begin{figure}[t!]
\centering
\includegraphics[scale=0.6]{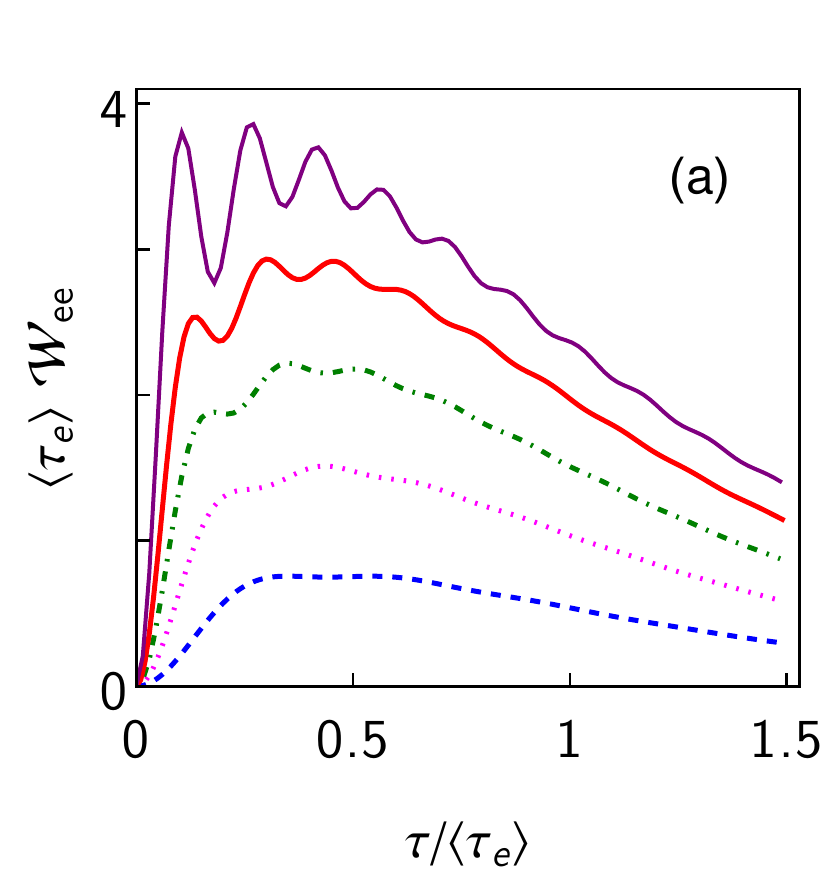}~~
\includegraphics[scale=0.6]{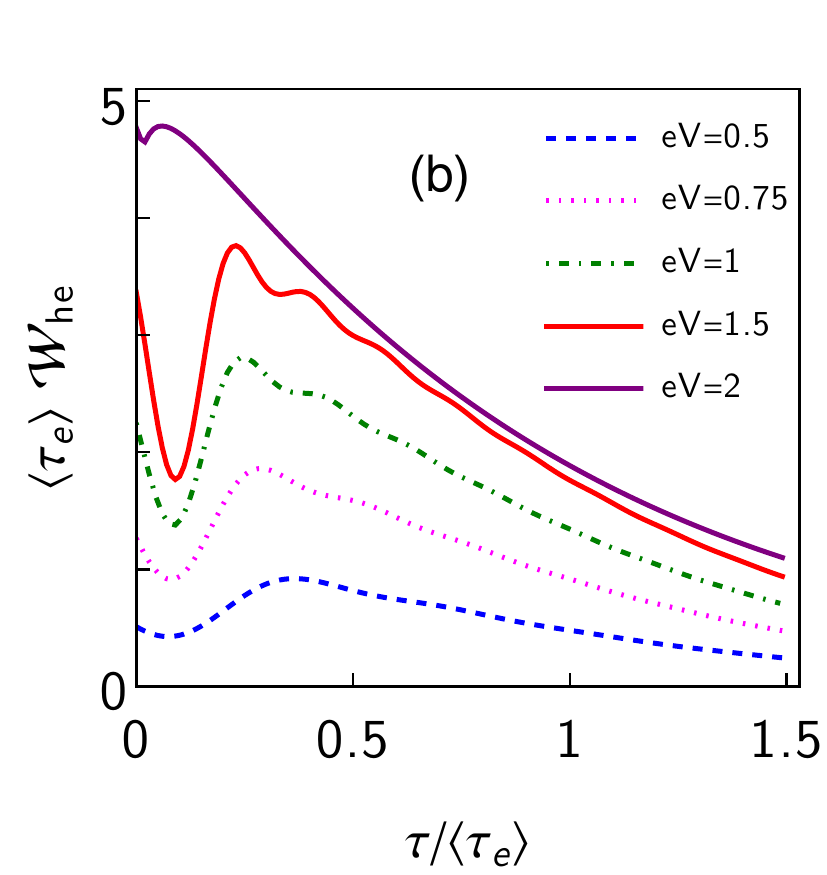}
	\caption{(a) $\langle \tau_{e}\rangle \mathcal{W}_{ee}$ and (b) $\langle \tau_{h}\rangle \mathcal{W}_{he}$ as functions of $\tau/\langle\tau_{e}\rangle $ at $\phi\!=\!\pi$ and $L_{\rm S}\!=\!1.5\xi$ for various energy windows $eV$. 
	In each figure, five curves from bottom to top are scaled by an integer $n \in\{1,2,3,4,5\}$ consecutively for the sake of clarity.}
\label{fig:Wee_Ls}
\end{figure}

\section{\label{sec:bias}Bias dependence of the waiting time distributions}
In the main text, we have shown results for $eV=\Delta/2$ only. To better understand the effect of the voltage bias, particularly for the parameter regime where the WTDs show anomalous behaviors, we now plot local $\mathcal{W}_{ee}$ and nonlocal $\mathcal{W}_{he}$ at $\phi\!=\!\pi$ for several transport windows. 
In \cref{fig:Wee_Ls}(a), $\mathcal{W}_{ee}$ shows a plateau region for $eV=\Delta/2$, which is exactly the same result shown in the main text for the same parameter values (just scaled with respect to the axis labels). 
Increasing the bias, small oscillations appear in the WTD which give rise to the plateau feature. The frequency and amplitude of the oscillations increase with $eV$, particularly for $eV\!\ge\!\Delta$, since, in addition to the zero energy anomaly from the MBSs, the transmission also features small peaks around $E\sim\Delta$, cf. \cref{fig:amplitude}(b). The period of these oscillations is related to the energy difference between peaks, that is, it is inversely proportional to $\Delta$~\cite{Haack2014,Mi2018,Schulz2022}. 

We have already described the anomalous behavior of $\mathcal{W}_{he}$ at $\phi=\pi$ in the main text. We now plot the nonlocal WTD with the same parameters for other bias values and concentrate on short waiting times $\tau$; longer waiting times are mostly unaffected by the bias voltage~\cite{Mi2018}. 
With increasing the bias, $\mathcal{W}_{he}(0)$ increases, see Ref.~[\onlinecite{Mi2018}], and so does the value of the dip or minimum in the nonlocal WTD. We also note an oscillatory behavior for larger biases. Therefore, the behavior of the WTDs for voltages larger than the one used in the main text, but that are still comparable to the superconducting gap, does not affect our main conclusions. 

\section{Nonlocal joint waiting time correlations\label{sec:joint-wtd}}
Finally, we show some results for the correlations between transmitted electrons and/or reflected holes to complete Table~I. 

\begin{figure}[!thpb]
	\centering
	\includegraphics[scale=0.38]{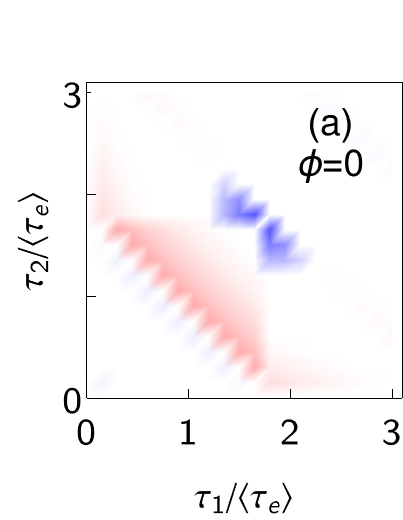}
	\includegraphics[scale=0.38]{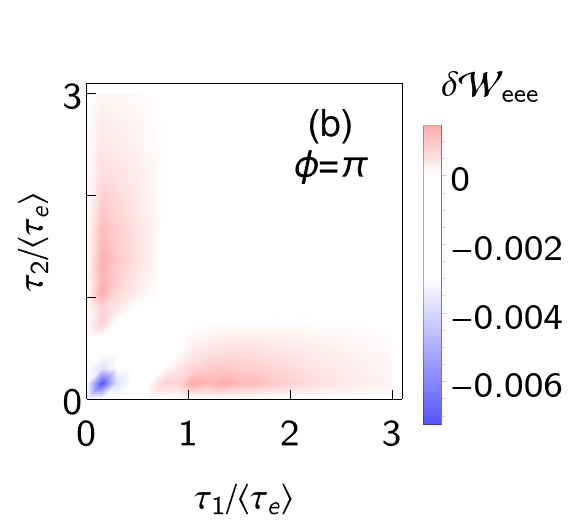}\\
	\includegraphics[scale=0.38]{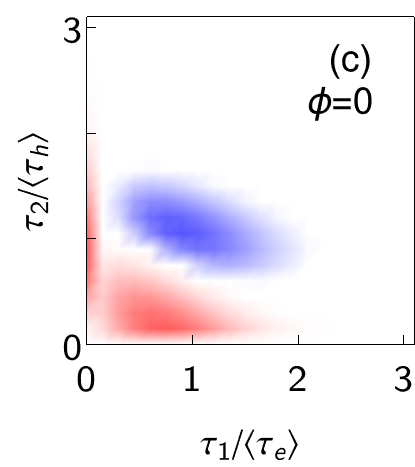}
	\includegraphics[scale=0.38]{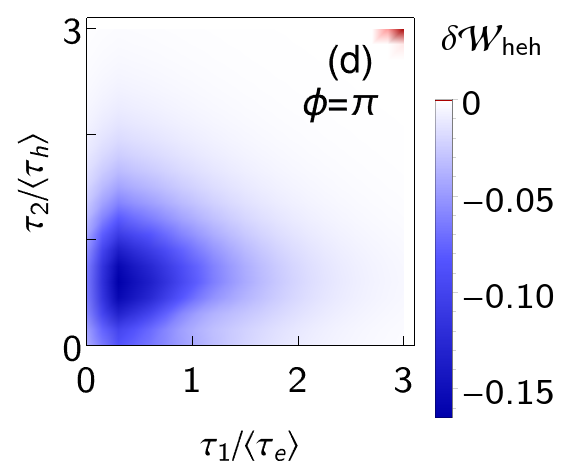}
	\caption{Correlations between waiting times for the detection of three consecutive electrons $\mathcal{W}_{\rm eee}$ and hole-electron-hole $\mathcal{W}_{\rm heh}$ for $L_{\rm S}\!=\!1.5$ as functions of $\tau_1$ and $\tau_2$ at $\phi=0$ (a,c) and $\phi=\pi$ (b,d).}
	\label{fig:Weee_eV}
\end{figure}

First, the sequential tunneling of three electrons into the right lead is almost uncorrelated, i.e., $\delta\mathcal{W}_{eee}(\tau_1,\tau_2)\simeq \mathcal{W}_{ee}(\tau_1)\mathcal{W}_{ee}(\tau_2)$, as seen in \cref{fig:Weee_eV}(a) and \cref{fig:Weee_eV}(b) for $\phi=0$ and $\phi=\pi$, respectively. 

Next, we show that the correlation between waiting times for the hole-electron-hole sequence of detections, i.e., $\delta\mathcal{W}_{heh}$, is not informative for the Majorana states. 
In the absence of any phase, there is a very small correlation between two nonlocal waiting times when both of them are at very short waiting times, i.e., $\tau_{1(2)} < \langle \tau_{e(h)} \rangle$, see \cref{fig:Weee_eV}(c). 
However, when $\phi=\pi$, the correlation dies out. Mostly negative values of the correlation exist indicating dominance of the uncorrelated part $\mathcal{W}_{he}(\tau_1)\mathcal{W}_{eh}(\tau_2)$, as observed in \cref{fig:Weee_eV}(d). 

\end{document}